\documentstyle[12pt]{article}
\input psfig.sty

\newcommand{\beq}{\begin{equation}}
\newcommand{\eeq}[1]{\label{#1} \end{equation}}
\newcommand{\insertplotlong}[1]{\centerline{\psfig{figure={#1},height=16.5cm}}}

\title{\bf Multi Module Model for Ultra-Relativistic Heavy Ion
Collisions}
\author{V. Magas$^{1,2,3}$\footnote{Talk given at the New Trend in High-Energy Physics, Yalta, Crimea,
Ukraine, September 22-29, 2001.}, L.P. Csernai$^{3,4}$ and D.
Strottman$^{5}$\medskip \\ {\it $^1$~CFIF, Instituto Superior Tecnico}\\
{\it Av. Rovisco Pais, 1049-001 Lisbon, Portugal}\\
{\it $^2$~Bogolyubov Institute for Theoretical Physics}\\
{\it Metrolohichna str. 14b, 01143 Kiev, Ukraine}\\
{\it $^3$~Section for Theoretical and Computational Physics}\\
{\it University of Bergen, Allegaten 55, 5007 Bergen, Norway}\\
{\it $^4$~KFKI Research Institute for Particle and Nuclear Physics}\\
{\it P.O.Box 49, 1525 Budapest, Hungary}\\
{\it $^5$~Theoretical Division, Los Alamos National
Laboratory}\\ {\it Los Alamos, NM, 87454, USA}
}

\begin{document}

\maketitle

\begin{abstract}
The Multi Module Model for Ultra-Relativistic Heavy Ion Collisions at RHIC
and LHC energies is presented.  It uses the Effective String Rope Model for
the calculation of the initial stages of the reaction; the output of this
model is used as the initial state for the subsequent one-fluid calculations.
It is shown that such an initial state leads to the creation of the third flow
component. The hydrodynamical evolution of the energy density distribution is
also presented.
\end{abstract}

%\section{Introduction}
%\label{ch-1}
A realistic and detailed description of an energetic heavy ion reaction
requires a Multi Module Model, wherein the different stages of the reaction are
each described with a suitable theoretical approach. It is important that
these Modules are coupled to each other correctly: on the interface, which is
a three dimensional hyper-surface in space-time, all conservation laws should
be satisfied and entropy should not decrease. These matching conditions were
worked out and studied for the correct matching at FO in detail in Refs.
\cite{FO1}.

Fluid dynamical models are widely used to describe ultra-relativistic heavy
ion collisions.  Their advantage is that one can vary flexibly the Equation of
State (EoS) of the matter and test its consequences on the reaction dynamics
and the outcome. For example, the only models which may handle the
supercooled QGP are hydrodynamical models with a corresponding EoS.  In
energetic collisions of large heavy ions, especially if a Quark-Gluon Plasma
(QGP) is formed in the collision, one-fluid dynamics is a valid and good
description for the intermediate stages of the reaction. Here, interactions
are strong and frequent, so that other models, (e.g. transport models, string
models, etc., that assume binary collisions, with free propagation of
constituents between collisions) have limited validity.

On the other hand, the initial and final, Freeze-Out (FO), stages of the
reaction are outside the domain of applicability of the fluid dynamical model.
After hadronization and FO, the matter is already dilute and can be
described well with kinetic models. The initial stages are the most
problematic.  Non of the theoretical models on the physics market can
unambiguously describe the initial stages (see for example
\cite{MCS01,Dr-th01} for discussion).

Thus, since we can not unambiguously describe the initial stages of
relativistic heavy ion collisions in microscopic models, we have to use some
phenomenological models.  There are two phenomenological models, which are
most frequently discussed in the literature - the Landau and the Bjorken
scenarios.  These models describe two extreme cases - complete stopping or
complete transparency, respectively.  Nowadays, experiments have entered the
region where the Bjorken model is expected to be applicable, but there is no
clear and unambiguous confirmation that experiments follow this scenario.  The
preliminary experimental results from RHIC do not show transparency - most
particle multiplicities (except maybe the most central collisions) do not
show a dip in the rapidity spectra, but rather a plateau around mid rapidity
\cite{QM01,multipl}, which is a sign of strong stopping. Furthermore, very
strong elliptic flow has been measured, which shows a clear peak around mid
rapidity \cite{QM01,v2}. To build such a strong elliptic flow, strong
stopping and momentum equilibration are required. Also the $\bar{p}/p$ ratio
at mid rapidity measured at RHIC \cite{QM-vid,prstar} (preliminary) is still
far from 1, which tells us that the middle region is not baryonfree.

%\section{Two Module Model}
%\label{MM2}
Our goal is to build a Multi Module Model for ultra-relativistic heavy ion
collisions valid for RHIC and LHC energies, and maybe for the most energetic
SPS collisions. In the present work only the first step is done - a Two Module
Model \cite{Dr-th01,CAMS01} - but a very important one, since the Effective
String Rope Model (ESRM) \cite{MCS01,Dr-th01} has been developed for the most
problematic module - the module describing the initial stages of collisions.

We describe the initial state with the ESRM; then the hydrodynamical
evolution with a QGP EoS is used for the intermediate state. The
hydrodymanical calculations are performed with the Los Alamos Particle-in-Cell
(PIC) one fluid code \cite{COS89,HYD}. The hydro evolution stops at the FO
hypersurface. We present a version of the code assuming that FO (simultaneous
chemical and thermal) happens on the simplified toothed hypersurface, where
it's normal vector, $d\sigma^\mu$, is parallel to the flow velocity for
every cell. On average this hypersurface approximates the constant time
hypersurface. Therefore the flow velocity does not change during the FO
process, and the calculations can be done in local rest frame of the matter.
Such a surface is also completely timelike; this allows us to avoid the
problems discussed in Refs. \cite{FO1}. 
The more advanced description of the
FO process is planned to be separated into the Third Module. 

The EoS presently used in the code was used and presented in Ref. \cite{COS89}.
This phenomenological phase transition model\\
A) takes the phenomenological EoS for hadronic matter in a simple form, which
nevertheless allows one to check
different parameterizations discussed in the literature;\\
B) uses the Bag model EoS to describe QGP;\\
C) creates a complete EoS, containing pure phases and a region
where they coexist, by the Maxwell construction.

\begin{figure}[t]
%\vspace{0.6cm}
\insertplotlong{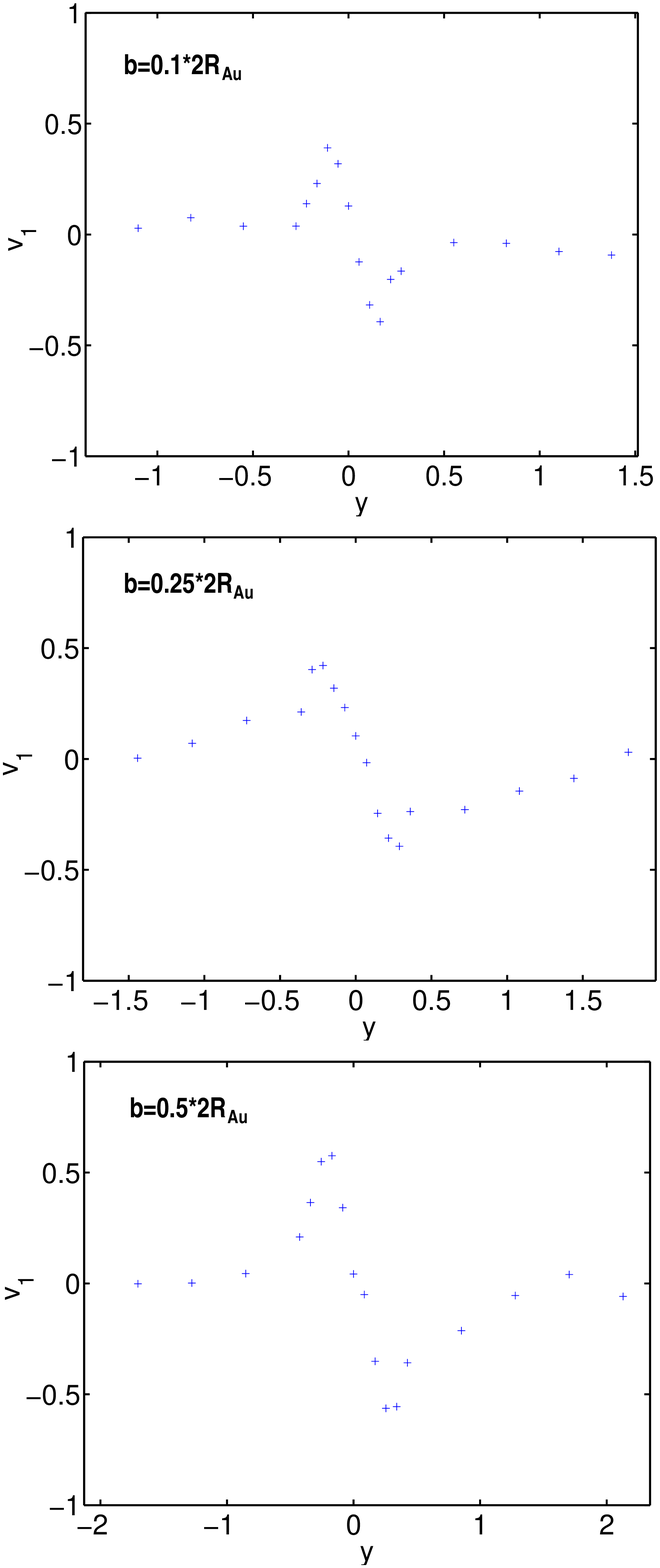}
\vspace{-0.5cm}
\caption{Preliminary results of Two Module Model calculations for
$v_1$ vs rapidity, $y$.
Au+Au collision at $\varepsilon_0=65\ GeV/nucl$ for different
impact parameters. Initial state was generated with $A=0.065$, $\tau=3.5\ fm/c$
(see \cite{MCS01,Dr-th01} for details).}
\label{v1y}
\end{figure}

The reader may notice that the model is still in a formative stage and a lot
of further work remains necessary. We are not yet ready to present the
quantitative calculations to be compared with data. Nevertheless we present
some preliminary results. First of all, the flow component, $v_1$, as a
function of rapidity, $y$, is presented in Fig. \ref{v1y}, to show that our
expectation that the model will generate a third flow component (Refs.
\cite{MCS01}) is reasonable \cite{Dr-th01,CAMS01}. Our initial state
generated by the ESRM indeed produces a strong antiflow in semi-central
collisions. The directed $v_1$ component appears to be very small, as
expected for RHIC energies
(for example \cite{v1}). 
The peaks in $v_1$
(actually $v_3$) around $y=\pm 0.3$ look very high, but this is typical for
calculations without thermal smearing \cite{15l}. Including thermal smearing
will lead to smaller and wider peaks.

\begin{figure}[t]
%\vspace{0.6cm}
\insertplotlong{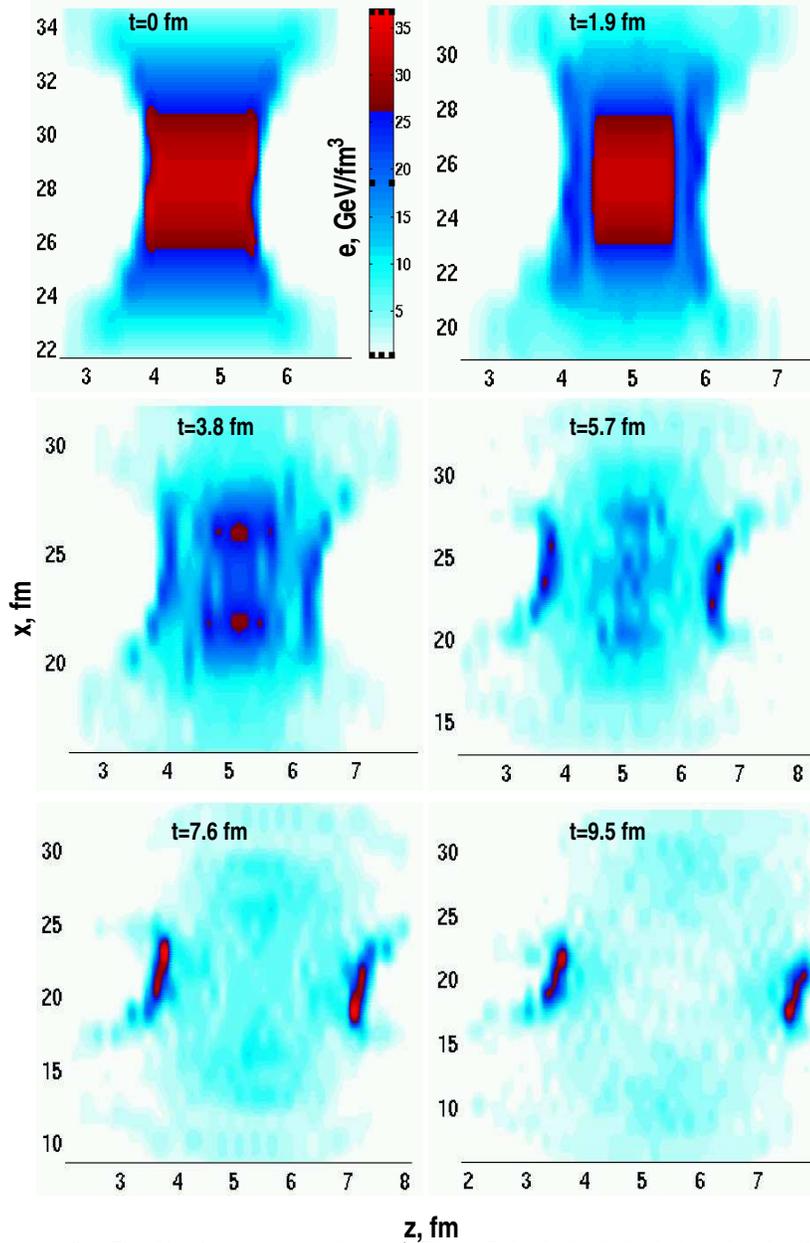}
\vspace{-0.5cm}
\caption{Preliminary results of Two Module Model calculations for
energy density distribution in the reaction plane.
Au+Au collision at $\varepsilon_0=65\ GeV/nucl$, $b=0.1\cdot 2R_{Au}$ 
for different
moments of time; initial state was generated with $A=0.08$, $\tau=3.5\ fm/c$
(see \cite{MCS01,Dr-th01} for details). }
\label{e-evol}
\end{figure}

We also present in Fig. \ref{e-evol} the evolution of the energy density
distribution, $e(r,t)$, during the hydrodynamic expansion (notice that $t=0$
corresponds to the moment then we start the hydro description, i.e.
$3.5\ fm/c$ after the collision). We can see the development of the small
but very dense, shock-like domains at the intermediate stages. This effect,
at least to some extent, is due to numerical artifacts. The PIC method has a
well-known defect called 'ringing' or marker-clustering \cite{ringing}. It is
called this way because early calculations in two dimensions resulted in
density distributions that looked similar to the vibration patterns in sound
vibrations - areas of higher and lower densities \cite{ringing}. 
%If one looks
%at the details, the large density in the fluctuations is limited to one cell
%(actually two because of reflection symmetry). If one eliminates these two
%cells, the remaining cells look very much alike in all calculations. 
Advanced
algorithms that mitigate this effect are being implemented in the code.

Another reason which may cause the development of the numerical artifacts is
a step-like in beam direction initial energy density distribution (output of
ESRM): it has a jump from $e(x,y)=const$ inside the matter to $0$ in the
outside vacuum. In order to avoid (or at least to suppress) the effect of this
we are planning to smooth over initial energy density distribution, for
example by a Gaussian shape, as proposed in Ref. \cite{hirano}. Thus
it is plausible that the development of the shock-like domains in the
simulation is not a physical effect, although we can not rule out other
possibilities.

To evaluate the observables a third, FO module is going to
be attached to the model. Here the widely used Cooper-Frye model should
be essentially improved and modified. First of all conservation laws and
the requirement of increasing entropy should be enforced in the module.
In addition, particularly for FO across space-like hypersurfaces,
realistic, non-equilibrium post FO phase space distributions
will have to be used to avoid negative contributions occuring in the
naive use of the Cooper-Frye model \cite{FO1,Dr-th01}.

\end{document}